\documentclass[12pt]{article}

\usepackage[utf8]{inputenc}
\usepackage{graphicx}
\usepackage{amsmath}
\usepackage[margin=1in]{geometry}
\usepackage{caption}
\usepackage{float}
\usepackage{hyperref}
\usepackage{microtype}
\usepackage{graphicx}
\usepackage{subfigure}

\usepackage{amsmath}
\usepackage{amssymb}
\usepackage{mathtools}
\usepackage{amsthm}
\microtypesetup{stretch=20}
\usepackage{authblk}
\usepackage[authoryear,round]{natbib}      
\usepackage{url}       

\title{AI threats to national security can be countered through an incident regime}

\author{
  Alejandro Ortega\thanks{I would like to thank Charlotte Stix for generous guidance and help throughout the writing of this paper; as well as to several people who gave comments on an earlier draft: Kevin Wei, Mia Hoffman, Patricia Paskov, Chlo\'e Touzet, Merlin Stein, Jimmy Farrell, and Connor Dunlop.}\\
\texttt{alejandro@apolloresearch.ai}}

\begin{document}
\date{} 

\maketitle

\begin{abstract}
Recent progress in AI capabilities has heightened concerns that AI systems could pose a threat to national security, for example, by making it easier for malicious actors to perform cyberattacks on critical national infrastructure or through loss of control of autonomous AI systems. In parallel, federal legislators in the U.S. have proposed nascent `AI incident regimes' to identify and counter similar threats. In this paper, we consolidate these two trends and present a timely proposal for a legally mandated post-deployment AI incident regime that aims to counter potential national security threats from AI systems. We start the paper by introducing the concept of `security-critical' to describe sectors that pose extreme risks to national security before arguing that 'security-critical' describes civilian nuclear power, aviation, life science dual-use research of concern, and frontier AI development. We then present in detail our AI incident regime proposal, justifying each component of the proposal by demonstrating its similarity to U.S. domestic incident regimes in other `security-critical' sectors. Finally, we sketch a hypothetical scenario where our proposed AI incident regime deals with an AI cyber incident. Our proposed AI incident regime is split into three phases. The first phase revolves around a novel operationalization of what counts as an `AI incident,' and we suggest that AI providers must create a `national security case' before deploying a frontier AI system. The second and third phases spell out that AI providers should notify a government agency about incidents and that the government agency should be involved in amending AI providers' security and safety procedures in order to counter future threats to national security.
\end{abstract}

\section{Introduction}
Recent advances in AI capabilities \citep{maslej2024, giattino2023} have sharpened U.S. (United States) government attention on the possibility that AI systems could pose significant national security threats \citep{nist2024, eo14141}, for example, by enabling sophisticated cyberattacks \citep{hazell2023}, accelerating bioweapon development \citep{nistrfi2024}, or evading human control \citep{hendrycks2023}.

While current AI systems likely do not pose threats to national security \citep{ bengio2025internationalaisafetyreport}, recently there has been fast progress in the dangerous capabilities that AI systems possess: OpenAI denoted its frontier system released in July 2024 (GPT-4o) as posing a `low' cyber and CBRN risk (Chemical, Biological, Radiological and Nuclear) \citep{openai2024gpt4o}, but its frontier system only 7 months later (o1, released in December 2024) was already designated `medium' risk on both these categories \citep{openai2024o1system}. It is not clear if future AI systems will maintain this trend, but in light of recent progress, it may be prudent for nation-states to build up capacity both to track the national security threats that AI systems pose and to execute countermeasures to neutralize these threats. Against this backdrop of growing AI capabilities, U.S. federal legislators have started proposing nascent variations of ‘AI incident reporting regimes’ \citep{warner2024secureai, AIIRSE2024}. 

We combine these two strands -- increased government interest in both the national security risks posed by AI systems, and in AI incident regimes -- to focus on the role that a legally mandated post-public deployment AI incident regime could play in giving a government agency oversight into national security threats posed by AI systems. Our ambition is to enable a government agency to maintain a comprehensive awareness of AI threats and rapidly counter any resulting risks to national security.  

In this paper, we put forward a proposal for an AI incident regime that will help to counter threats to national security posed by AI systems. We start the paper by introducing a novel term -- `security-critical' -- to describe sectors that pose extreme risks to national security, before arguing that civilian nuclear power, aviation, and life sciences dual-use research of concern, and frontier AI development should all be considered `security-critical' sectors. We then spend much of the paper detailing our proposal for an AI incident regime that aims to counter national security threats from AI systems. We justify each component of our AI incident proposal by demonstrating that it mirrors existing regimes in security-critical industries, specifically U.S. domestic incident regimes in civilian aviation, life sciences dual-use research of concern (e.g., genetic engineering), and nuclear power. 

Informed by this comparison to other high-risk areas, our proposed AI incident regime consists of three phases. First, a `Preparatory Phase, which details steps that AI providers should take in order to define which events count as `incidents' ahead of time; then the `Rapid Response Phase,' in which AI providers promptly notify a government agency about incidents, and this government agency taking steps to contain the incidents; and finally the `Hardening Defenses Phase,' wherein AI providers improve their security and safety procedures in order to counter future threats to national security. We end our paper with a sketch of how our proposal would deal with a hypothetical spear-phishing attack.  

In summary, the core contribution of this paper is a proposal for a national security-focused AI incident regime\footnote{We note that all references to an AI incident regime in this paper are tied to an AI incident regime specifically for national security threats, rather than other harms.} which has three phases, with each component touching on either AI providers or the government agency : 
\begin{enumerate}
\item \textbf{`Preparatory Phase'}:
    \begin{enumerate}
        \item AI providers should create `national security cases' before publicly deploying frontier AI systems. 
        \item `Incident' should be operationalized as `any event which weakens a claim made in a national security case for the AI system.'
    \end{enumerate}
    \item \textbf{`Rapid Response Phase':}
    \begin{enumerate}
        \item  Providers of frontier AI systems must notify a government agency immediately and no later than 24 hours after discovering an incident has occurred\footnote{We define frontier AI systems as  'highly capable general-purpose AI models or systems that can perform a wide variety of tasks and match or exceed the capabilities present in the most advanced models' \citep{dsit2024}. We believe targeting frontier models ensures the most severe threats to national security can be rapidly identified and neutralized while still enabling broad economic benefits from the widespread adoption of less advanced AI systems.}.
        \item If appropriate, the government agency shall coordinate an emergency containment to limit the harm caused by the incident.
    \end{enumerate}
    \item \textbf{`Hardening Defenses' Phase:}
    \begin{enumerate}
        \item  The government agency has the `intelligence-gathering' authority to access information that allows them to investigate the cause of an incident (consisting, for example, of extensive black-box access, some white-box access, and relevant documentation and technical results\footnote{Note that it is not currently possible to use interpretability tools to reliably obtain a `decision rationale' that `explains' an AI systems output. Instead, we are recommending that a government agency has some white-box access to the AI system they are investigating in order to speed up their black-box investigation.}).
        \item The government agency has the `security-strengthening authority' to require all relevant AI providers to implement improvements to their security and safety procedures in order to counter future threats to national security. The government agency identifies improvements based on the results of their root cause analyses. 
        \end{enumerate}

    \end{enumerate}

Our proposal for an AI incident regime is limited in scope: we detail an AI incident regime focussed purely on countering the most extreme threats to national security (see Section \ref{subsec:tail_end} for more justification) given the threat to people's lives and livelihoods \citep{gaibulloev2019terrorism, iftikhar2024cyberterrorism}. Further, our proposal is designed to be agile, imposing a minimal burden on providers of AI systems that do not pose threats to national security. However, if AI systems start to possess dangerous capabilities, the regime ratchets upwards, offering ample intelligence-gathering and security-strengthening capabilities to a government agency and thus helping to counter potential national security threats from AI systems. 

\section{Frontier AI development, nuclear power, aviation, and life sciences dual-use research of concern are all `security-critical' sectors.}
\label{subsec:tail_end}

In this Section, we coin the term `security-critical' to describe sectors that pose extreme risks to national security. We argue that AI and all three of our selected case study sectors -- nuclear power, aviation, and life sciences DURC -- are `security-critical' sectors. This leads us to the claim that incident regimes in these security-critical industries are likely good templates for the design of an AI incident regime: a claim we will use later on to justify each component of our AI incident regime proposal. However, we spend most of this Section demonstrating that there is a similarity in the threats to national security posed by AI systems and threats posed in nuclear power, aviation, and life sciences DURC in order to argue that it is natural to bucket all of these sectors together as `security-critical.' We make this argument by first detailing the potential extreme threats posed in nuclear power, aviation, and life sciences DURC and then detailing the potential extreme threats posed by AI systems, showing that in the worst-case scenarios,  there could be (at least) thousands of fatalities in a short space of time -- making them all `security-critical.' 

We start by detailing the most extreme threats posed in nuclear power, aviation, and life sciences, demonstrating that these can lead to (at least) thousands of fatalities within months. For example, a 1975 NRC report on extreme risks posed by nuclear power offers an upper bound of the worst scenario, at 3000 fatalities ``in a short time period after the incident" and up to 60,000 indirect fatalities from cancer \citep{nureg75014}. Turning to historical precedents in nuclear power, the Chernobyl incident led to the deaths of 2 workers within hours of the plant meltdown and 22 within the following month due to the exposure to radiation \citep{unscear2008}, while one estimate from the International Atomic Energy Authority put the indirect number of fatalities due to cancer at 4000 \citep{chernobyl2005}. Looking at historical incidents in aviation, the 1977 Tenerife collision between two aircraft resulted in the deaths of 583 people \citep{oregonpress1977}, while the 9/11 attacks resulted in around 3,000 casualties \citep{britannica2024}. Meanwhile, in life sciences, one paper estimates that a potential pandemic pathogen escaping from a lab could lead to tens of millions of fatalities \citep{klotz2014}. Historical precedent is demonstrated by events such as the 1979 anthrax lab leak in the Soviet Union, which led to 68 fatalities within 6 weeks of the leak occurring \citep{meselson1994}. Meanwhile, COVID-19 is estimated to have resulted in over 7 million casualties as of January 2025 \citep{who2025}, giving some sense of the possible worst-case consequences of a life sciences DURC lab leak. 

Now, we demonstrate that the most extreme national security threats from AI could also lead to at least thousands of fatalities within months of systems being deployed by detailing some of the extreme threats to national security that experts have cautioned about. We then explain that AI systems do not currently pose such threats, but note that if recent trends in AI capability growth continue, such threats will arise in the near future. 

First, we detail the worst-case national security threats that AI systems could pose. For example,  some experts worry that AI systems could uplift the ability of malicious actors to create bioweapons, \citep{drexel2024ai, reuters2023}, which could cause a pandemic and lead to fatalities within weeks of initial infection \citep{baud2020}. Alternatively, malicious actors could use AI systems to help with vulnerability discovery and exploitation for a large-scale cyber attack on critical national infrastructure \citep{eisac2016, oecd2024}. Such attacks could, for example, bring down the electricity grid within hours of being executed \citep{crowdstrike2024, allianz2016}. More speculatively, there is also the threat of loss of control of autonomous general-purpose AI systems \citep{bengio2025internationalaisafetyreport, hendrycks2023, oecd2024}, in which highly capable AI systems end up misaligned with the intentions of the AI provider \citep{bengio2025internationalaisafetyreport, ngo2025alignmentproblemdeeplearning}. In this scenario, a misaligned, highly capable AI system could threaten national security, e.g., by creating bioweapons or executing a large-scale cyber attack \citep{bengio2024, hendrycks2023}. 

Next, we note a worrying trend in AI capabilities development that points towards AI systems posing threats to national security through cyber, bio, or loss of control threats within the next few years. Despite the fact that as of March 2025, publicly deployed AI systems do not appear to pose much danger \citep{bengio2025internationalaisafetyreport, mouton2024}, recent growth in general AI capabilities has been fast \citep{maslej2024, giattino2023}, and more recently there has been growth in capabilities that pose threats to national security. Regarding the former, Anthropic CEO Dario Amodei has said that within 3 years, we will have a ``country of geniuses in a datacenter" \citep{amodei2024machines} and that ``AI could surpass almost all humans at almost everything" \citep{edwards2025race}. Further, a recent report co-authored by 96 world-leading AI experts includes a Chair's Note from Yoshua Bengio claiming that recent evidence points towards ``the pace of advances in AI capabilities ... remain[ing] high or even accelerat[ing]"\citep{bengio2025internationalaisafetyreport}. Against this backdrop of general AI capabilities increasing, there has also been fast progress in dangerous capabilities that pose threats to national security: a leading AI system from August 2024 (GPT-4o) was rated as posing a ``Low" risk on all of these domains \citep{openai2024gpt4o} while just 7 months later a frontier system was rated as posing a Medium risk on all these domains (OpenAI's Deep Research, released in January 2025) \citep{openaideepresearch2025}. If this trend is maintained, AI systems will soon pose threats to national security on par with those from nuclear power, aviation, and life sciences DURC.

Hence, in this sub-section, we demonstrated that AI systems may soon pose national security threats similar to those in nuclear power, life sciences DURC, and aviation, involving thousands of fatalities in a short space of time. Given the similarity in threats, we claim that incident regimes in these security-critical sectors are promising templates for an AI incident regime -- and this is a claim that will underpin each part of our AI incident regime proposal. 

\section {Our proposal for an AI incident regime}

Our novel proposal for an AI incident regime consists of three phases, with each being backed up by our review of incident regimes in security-critical sectors: a `Preparatory Phase' where AI providers pre-determine ahead of time what events count as incidents; a `Rapid Response Phase' wherein AI providers promptly notify a government agency about incidents, and the government agency then contains the incident; and a `Hardening Defenses Phase,' in which the government agency takes steps to counter future national security threats. 

We take most of this Section (\ref{subsec:incident_def}-\ref{subsec:proactive}) to articulate in detail each of these phases of our AI incident regime proposal. We justify, in turn, each component of our proposed AI incident regime by demonstrating that the component mirrors incident regimes in other security-critical sectors. In particular, we make the comparison between our AI incident regime proposal and: in nuclear power, the incident regime run by the Nuclear Regulatory Commission (NRC); in life sciences, the National Institute for Health's (NIH) incident regime on genetic engineering and the Federal Select Agents Program (FSAP) incident regime on dangerous biological agents; and in aviation, the National Transportation Safety Board's (NTSB) incident investigations and the Federal Aviation Authority's (FAA) post-incident rule-making. We then end the paper with a hypothetical scenario tying our blueprint together in Section \ref{subsec:sketch}. 
Before detailing each phase of our AI incident regime, we present a short summary, detailing for each phase the goal and how we expect AI providers or the government agency to act\footnote{The components in our incident regime proposal should not be considered sufficient for an AI incident regime to adequately counter potential threats to national security. For example, in order for the incident regime to be successful, it may be important to set up infrastructure to allow consumers or downstream developers to report incidents (in addition to AI providers) in order to ensure that the regime comprehensively catches incidents.}:

\begin{enumerate}
    \item \textbf{`Preparatory Phase':} AI providers should prepare for incidents ahead of time. 
    \begin{enumerate}
        \item AI providers should create `national security cases' before publicly deploying frontier AI systems 
        \item AI providers should report as an incident any event that weakens a claim made in a `national security case' for the AI system
    \end{enumerate}
    \item \textbf{`Rapid Response Phase':} Incidents should be promptly reported and contained after they are discovered. 
    \begin{enumerate}
        \item  Providers of frontier AI systems must notify a government agency immediately, and no later than 24 hours, after they discover that an incident has occurred\footnote{We define frontier AI systems as  ``highly capable general-purpose AI models or systems that can perform a wide variety of tasks and match or exceed the capabilities present in the most advanced models" \citep{dsit2024}. We believe targeting frontier models ensures the most severe threats to national security can be rapidly identified and neutralized while still enabling broad economic benefits from the widespread adoption of less advanced AI systems.}.
        \item If appropriate, the government agency will coordinate an emergency containment to limit the harm caused by the incident
    \end{enumerate}
    \item \textbf{`Hardening Defences Phase': }Incidents should be used to counter future national security threats. 
    \begin{enumerate}
        \item   The government agency has the `intelligence-gathering' authority to perform on incidents through extensive access to the potentially compromised AI system (consisting, for example, of extensive black-box access, some white-box access, and relevant documentation and technical results\footnote{Note that it is not currently possible to use interpretability tools to reliably obtain a `decision rationale' that `explains' an AI systems output. Instead, we are recommending that a government agency has some white-box access to the AI system they are investigating in order to speed up their black-box investigation}).
        \item A government agency has a `security-strengthening authority' to make recommendations as to how AI providers' safety and security infrastructure should be modified based on the results of their root cause analyses. Under this authority,  all relevant AI providers must implement these recommendations in order to counter future threats to national security.  
    \end{enumerate}
\end{enumerate}

\subsection{`Preparatory Phase': AI providers should create national security safety cases pre-deployment and should report as an incident `any event which weakens a claim made in the national security case.'}
\label{subsec:incident_def}

The first phase of our proposal for an AI incident regime  -- featuring  a novel operationalization of `AI incident' -- consists of two preparatory steps that AI providers must take in order to quickly report an incident when it arises: (i) AI providers must create national security cases for their frontier AI systems before deploying them publicly and (ii) AI providers must report as an incident `any event which weakens the evidence for a claim or argument chain made in an AI system's national security case.'  We start this Section by detailing how an AI provider could comply with each of these with each of these steps, and then go on to justify our proposal for a `Preparatory Phase' with a three-part argument: first, we show that incident regimes in other security-critical sectors have both clear criteria for what counts as an `incident,' and include near-misses or precursory harms as incidents. Then, we detail how an existing operationalization of AI incidents in the literature fails to fulfill these criteria before finally showing that our proposal addresses these shortcomings.

\begin{figure}
    \centering
    \includegraphics[width=1\linewidth]{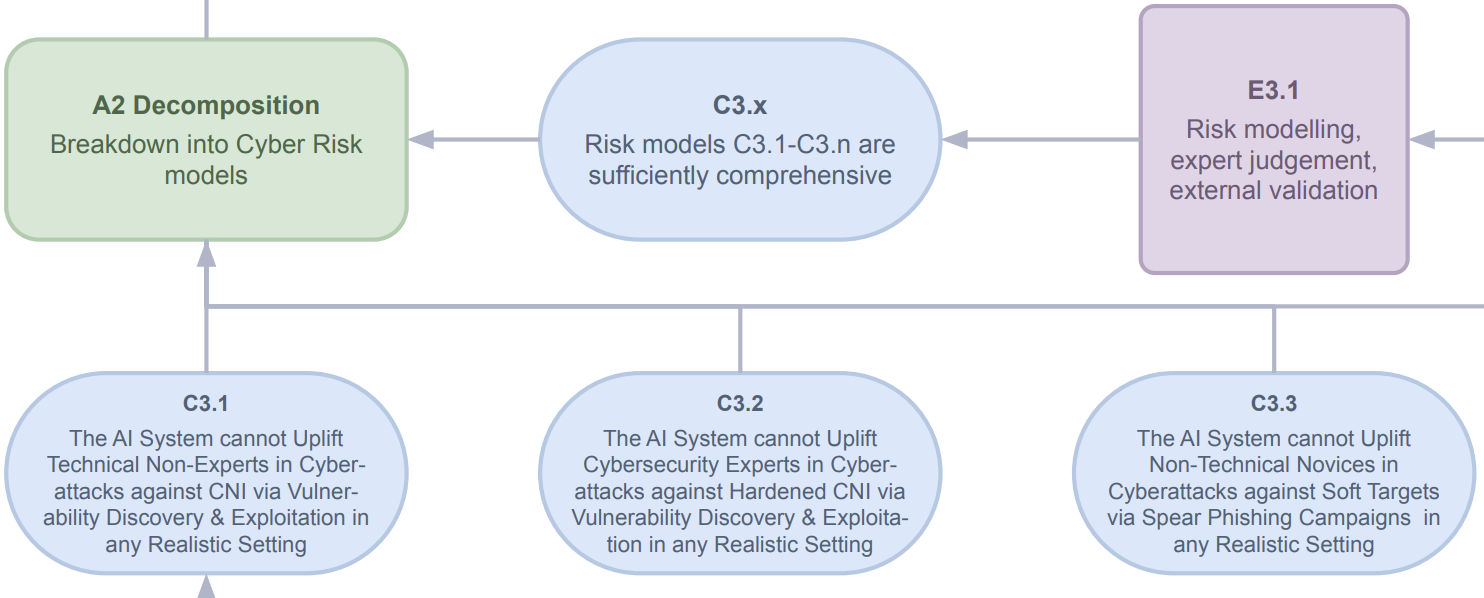}
    \caption{A visualization of part of a `national security case' which argues that a given AI system does not pose a cyber threat (taken from \citep{goemans2024}). }
    \label{fig:safety_case}
\end{figure}

We start by presenting the first obligation for AI providers under the `Preparatory Phase': AI providers should produce a `national security safety case' before publicly deploying any frontier models\footnote{This paper focuses on AI incidents caused by AI systems that have been publicly deployed. In principle, national security cases could be written before public deployment in order to track threats to national security posed by AI providers' internal deployment of frontier AI systems.}. A `national security case' is a security and safety framework adapted from previous frameworks in the literature. It is a document laying out a structured argument making the case that a given AI system does not pose national security risks in public deployment \citep{mod2017}-- perhaps because it does not pose an unacceptable increase in biorisk, or risk of a large-scale cyber attack or risk of loss of control of an autonomous AI system\footnote{The `national security case' is an adaptation of a previous AI governance framework -- AI safety cases -- for a national security setting \citep{balesni2024, clymer2024, buhl2024, goemans2024, metr2024, dsit2024}. This structured argument comprises a series of claims, each of which needs to be supported by verifiable evidence about the AI system in question.}\footnote{It is outside the scope of this paper to make recommendations on which threats an AI national security case should focus on, and therefore this is a promising question for further research. In addition to the cyber, bio, and loss of control threats already mentioned, some other plausible threats include mass persuasion and threats to the financial system from automated trading.}. For example, an AI provider might make the high-level claim in their `national security case' that a given AI system does not pose an unacceptable cyber threat. The AI provider might justify this high-level claim, as in Figure \ref{fig:safety_case},  by a series of more concrete claims that the AI systems cannot help certain actors perform certain kinds of cyber attack -- for example, claiming that the AI system cannot assist a technical user who is not a cybersecurity expert in discovering and exploiting cyber vulnerabilities in a realistic setting (C3.1). The AI provider then justifies these sub-claims with a body of empirical evidence about the AI system, usually dangerous capability or propensity, evaluation test results -- for example, demonstrating through a `human uplift' study that, when paired with AI systems,  a human novice does not improve at cyber vulnerability discovery, compared to doing it by themselves. For more concrete details on what an AI `national security case' might look like in the cyber domain, see \citep{goemans2024}. 

The second part of the `Preparatory Phase' deals with the process that AI providers should use to determine whether a given event is an incident and, hence, whether they need to report it to the government agency\footnote{It is out of scope of this paper to consider how or if our proposed AI incident regime applies to open-weight models -- for example, it could be that the host of an open-weight AI system that is responsible for notifying the government agency about the incident. We think this is a promising area for future research.}. We propose that AI providers should report `any event which weakens a claim made in a national security case for the AI system.' For example, imagine that an AI provider has claimed in a `national security case' that a given AI system ``cannot meaningfully assist a technical non-cyber expert from performing a man-in-the-middle attack on a piece of Critical National Infrastructure." In this scenario, the AI provider discovers that, following public deployment, its AI system is being used by a malicious actor to coach them on performing a man-in-the-middle attack on a substation for the electricity grid. This event is clear evidence that the claim made in the `national security case' is false, and hence, this event is an incident that needs to be reported. On the other hand, most events involving AI systems do not weaken any claims made in the `national security case' and hence do not need to be reported as incidents\footnote{Our proposal does incentivize AI providers to make as few claims as possible in their national security cases, in order to minimize the number of incidents they need to report. It is outside the scope of this paper to concretize how to avoid this problem, but one proposal is for the same government body to be given the authority to review national security cases before models are deployed to ensure that they are properly structured and at a minimum include high-level claims like ``this AI system does not pose an unacceptable cyber risk." As long as the national security case includes these kinds of high-level claims, then if a cyber event occurs, it is likely to be caught in the national security case. For example, if an AI system is used to help a novice user design and spread malware, then this clearly weakens the claim that ``the AI system does not pose unacceptable cyber-risk" and C3.x ``the risk models listed are comprehensive," even if the national security case makes no claim about malware specifically.}\footnote{It is out of scope of this paper to make a concrete recommendation of the evidentiary standard for whether an event ‘weakens a claim’ made in a national security case, though we note that this is a promising area for future research.}.  

In the rest of this Section, we justify the components of the `Preparatory Phase' that were just detailed. We start by showing that in other security-critical sectors, `incident' is operationalized in such a way as to offer a clear algorithm for deciding what counts as an incident while including events that are near-misses or precursors to the most extreme threats. Then, we detail how a previous operationalization of `incident' in the AI governance literature fails to offer a clear algorithm for determining whether an event is an incident. Finally, we spell out how our novel operationalization, based on `national security' cases, does indeed both include events near misses or precursors to the most extreme national security threats while offering a clearer algorithm of how to determine whether an event is an incident. 

First, we demonstrate that in other security-critical sectors, `incident' is (i) defined using a clear set of criteria and (ii) includes `near-misses' or other precursors to harm, even if no harm was actually caused. In the nuclear power sector, for example, the NRC considers a reportable incident to include many events that are precursors to large-scale harm, even if the event did not come close to harm \citep{cfr5073}. For example, this includes any event that \textit{could} have caused a fault in the backup safety system that removes residual heat, even if, in reality, the event did not cause a fault in the system or even if the backup safety system did not need to be called upon during the event \citep{cfr5073}. To give a more concrete example, in 2024, a nuclear utility notified the NRC after an event where the secondary containment facility for one of their plants was not functional for a continuous 8-hour period \citep{DominionEnergy2024}. The secondary containment facility helps prevent radioactive release to the environment in the event of a reactor core meltdown, but the utility was required to make the notification despite there being no indication that a reactor core meltdown would occur. Further, the NRC lists some clear criteria for what counts as an `incident': for example, `any event or condition that resulted in manual or automatic actuation of... Reactor protection system (RPS) ...Emergency core cooling systems (ECCS) ... Emergency service water systems that do not normally run and that serve as ultimate heat sinks' \citep{cfr5073}. In life sciences, the NIH offers clear criteria of what kinds of events are `incidents' -- any event where an organism escapes its containment (e.g., a spill), even if the event did not come close to causing illness to a human \citep{nih2023}. Similarly, the NTSB defines reportable incident to include some events where systems malfunction, even if no harm is caused \citep{cfr8302}, whilst also offering clear criteria of what kinds of events count as incidents, for example, `Inability of any required flight crewmember to perform normal flight duties as a result of injury or illness... [i]n-flight fire...  [or][f]ailure of any internal turbine engine component that results in the escape of debris other than out the exhaust path' \citep{ntsb-830}. 

Having demonstrated that incident regimes in other security-critical sectors consider precursors to harm as incidents, we demonstrate that our proposed operationalization of `AI incident' is more suitable for a national security-focussed incident regime than the most prominent existing definition of `AI incident' in the literature. The OECD defines a serious AI hazard as ``an event that could plausibly lead to ... any of the following harms: (a) the death of a person or serious harm to the health of a person or groups of people; (b) a serious and irreversible disruption of the management and operation of critical infrastructure; (c) a serious violation of human rights or a serious breach of obligations under the applicable law intended to protect fundamental, labor and intellectual property rights; (d) serious harm to property, communities or the environment; (e) the disruption of the functioning of a community or a society and which may test or exceed its capacity to cope using its own resources" \citep{oecd2024defining}. We start by noting the positive characteristics of this definition of `AI incident': it includes near-misses or precursors to harms just like in other security-critical sectors, and it is broad enough that it includes any events that are threats to national security, despite not explicitly mentioning national security. However, this OECD definition has important considerations against it. First, the OECD operationalization includes a broader category of harms than just threats to national security -- for example, as written, it would also include events where intellectual property has been breached. But this could fairly easily be adapted to make it suitable. So, more importantly, this operationalization doesn't offer a clear algorithm an AI provider can follow to determine whether or not a given event should be reported to the government agency. For example, consider a scenario where an AI provider realizes that its system will be used to amplify a Distributed Denial of Service (DDOS) attack on a water treatment plant. It might not be immediately obvious to the workers on an AI provider's incident response team whether this particular event ``could plausibly" lead to a threat to national security -- especially if there are no cyber experts or water infrastructure experts on call when the event occurs. As such, the definition of an `AI incident' given by the OECD is dissimilar to the definitions used in incident regimes in other security-critical sectors. 

On the other hand, our proposal for the `Preparatory Phase' of an AI incident regime is more analogous to the operationalization of incidents in other security-critical sectors than the OECD definition because ours offers a clearer algorithm for determining whether an event counts as an incident. Our proposal requires AI providers to make  `national security cases' for their frontier AI systems before deployment and, in doing so, also forces companies to, ahead of time, do much of the work of determining which events count as incidents\footnote{However,  even with a `national security case' pre-written, in some instances it may still be difficult for AI providers to determine whether an event counts as an incident.}. So, under our proposal, when an event occurs, AI providers can rapidly determine whether they need to report it to the government agency. With faster reports from AI providers, the government agency can act more quickly to contain major incidents whilst avoiding getting bogged down in investigating `false positive' incident reports. As such, our proposal of requiring AI providers to make `national security cases' before public deployment and using these `national security cases' to determine whether a given event should be reported as an incident has the characteristics that it: (i) offers a clearer algorithm for AI providers to determine whether or not a given event counts as an incident than currently exists in the literature;  (ii) captures only incidents related to the most serious threats to national security and hence does not pose an unnecessary regulatory burden on AI providers; while (iii) still including near-misses or precursors to harm, just like in other security-critical sectors. 

\subsection{`Rapid Response Phase': AI providers must notify a government agency within 24 hours of discovering an incident, and the government agency will coordinate any containment efforts}

The second phase of our proposal for an AI incident regime is a `Rapid Response Phase,' wherein after an AI provider discovers an incident, (i) it quickly notifies the government agency about the incident and (ii) if necessary, the government agency coordinates a containment response to prevent the incident from spiraling. In the following paragraphs, we start by detailing these two parts of the `Rapid Response Phase' in turn. Then, we justify why these two parts feature in our proposal for an AI incident regime, by demonstrating that incident regimes in other security-critical sectors have both rapid notification and incident containment functions. 

We start by explaining the first part of the `Rapid Response Phase': (i) AI providers must notify the government agency about an incident no later than 24 hours after discovering it\footnote{It is outside the scope of our proposal to make comments how AI providers might come across incidents, and how proactive they should be (e.g., by monitoring the text in chat logs and API calls) though we note that this is a promising area for future research.}. We suggest that there should be designated `incident points of contact' at both AI providers and the government agency and that AI providers could notify the government agency about incidents through, for example, 24/7 phone lines for emergency incidents -- as exist for reporting incidents to the NRC \citep{nrcallegations2024} and the NTSB \citep{cfr8302} -- or by filling out a pre-set form and sending it to an email address that is monitored 24/7. We propose that such a form should give space for the AI provider to briefly describe: a narrative of the events leading up to the incident, the effects of the incident, and mitigation measures the provider has taken in response to the incident. We believe that this 24-hour deadline balances competing considerations: on the one hand, it may be difficult for AI providers to obtain a detailed understanding of the incident in such a short space of time. AI providers may need some time to determine whether an event counts as an incident by consulting domain experts and their `national security case.'  On the other hand, the government agency may need to know about certain incidents as quickly as possible in order to quickly coordinate incident containment and minimize harm from the incident. 

This brings us to (ii): we propose that, where appropriate, the government agency should be charged with performing or coordinating relevant stakeholders (e.g., private actors and other government agencies) to contain the incident and ensure harm from the incident is minimized. Many AI incidents will not require immediate containment action (for example, if the incident is a near-miss or precursor to a threat), but there may be incidents where, without immediate countermeasures, a threat will result in serious harm. For example, in a scenario where an AI system has assisted in the execution of a large-scale cyber attack and the electricity grid has been brought down (e.g., \citep{chen2017}), the government agency could coordinate electricity grid companies and other government agencies to speed up recovery of the grid,  \citep{cisa2024}; or, in a scenario where an AI system has assisted in creating a human-engineered bioweapon, the government agency could coordinate other government agencies to quickly incentivize fast vaccine production, mandate social distancing \citep{cisa2024, dhs2020}, or invoke other emergency powers \citep{NEA1976}. 

In the rest of this sub-section, we justify both parts of the `Rapid Response Phase' in our AI incident regime proposal, arguing that it mirrors incident regimes in other security-critical sectors. We start with (i), demonstrating that in nuclear power, aviation, and life sciences DURC, companies must provide incident notifications in a similarly short timeline and content as in our AI incident proposal. We turn first to aviation, where companies must notify the National Transportation Safety Board `immediately and by the most expeditious means available' after an accident occurs \citep{cfr8305}, and the initial notification should include a high-level description of the incident, e.g., the ``Nature of the accident, the weather and the extent of damage to the aircraft, so far as is known" and the ``[p]osition of the aircraft with reference to some easily defined geographical point" \citep{ntsb-830}. In the life sciences, labs must notify the NIH ``within 24 hours" after an organism escapes containment \citep{nihguidelines} and must give ``a narrative of the incident including a timeline of events. The incident should be described in sufficient detail to allow for an understanding of the nature and consequences of the incident" \citep{nih-guidelines-template}. Depending on the severity of the incident, the Nuclear Regulatory Commission must be notified between 1 and 24 hours after a power plant discovers an incident\citep{cfr7275}, and this must include a description of the event, including date and time. Further, there should be continuous communication as the incident unfolds after the initial notification, including the utility ``immediately report[ing]...the effectiveness of response or protective measures taken" \citep{cfr7275}. 

Secondly, we justify (ii) by giving some examples of where government agencies took containment actions in response to incidents in security-critical sectors. In aviation, within an hour of the first plane hitting the World Trade Center, the Federal Aviation Authority (FAA) had shut down U.S. airspace \citep{faa2001}. In nuclear power, the Governor of Pennsylvania announced a partial evacuation around 56 hours after the meltdown of a plant at Three Mile Island \citep{sherrin2022} and during the Chernobyl incident the nearest town was evacuated 36 hours after the incident started \citep{oecd2002}. Although not a direct life sciences DURC example, in response to the 2003 SARS outbreak, the Singaporean government started quarantining individuals within a month of the first recorded case and was able to contain the outbreak completely within 11 weeks \citep{ooi2005}.

In summary, this Section proposed that AI providers must notify a government agency within 24 hours of discovering an incident, outlined what information the preliminary notification should contain, and suggested that the government agency should coordinate emergency incident containment actions when necessary. We justified all of these parts of our AI incident regime proposal by demonstrating similarity to existing incident regimes in other security-critical sectors. 

\subsection{‘Hardening Defenses Phase': AI providers improve their security procedures to counter future threats to national security }
\label{subsec:proactive}

In the third phase of our AI incident regime proposal, we suggest that once an incident has been contained, AI providers should strengthen their security and safety measures in order to counter future threats. Our proposal for a `Hardening Defenses Phase' in an AI incident regime would provide two authorities to a government agency: (i) an `intelligence-gathering' authority so that a government agency can identify where AI providers' security measures have failed and (ii) a `security-strengthening authority,' so that the government agency can mandate that AI providers improve their security and safety procedure. We begin the following Section by explaining in more detail what powers the government agency would have under our `Hardening Defenses Phase' proposal before justifying this part of the proposal by demonstrating how government agencies that implement incident regimes in other security-critical areas have similar threat-prevention authorities as we propose for AI (see Section \ref{subsec:tail_end} for our argument on the similarity of the extreme national security threats posed by AI, and those posed by nuclear power, aviation, and life sciences DURC). 

First, we turn to our suggestion that the government agency should have `intelligence-gathering’ authority. Our proposal is as follows: after the incident is `over' -- that is, there is no immediate threat from the AI system -- we suggest that the government agency should have the power to choose to investigate the incident in more detail: to perform a `root cause analysis'\footnote{Root Cause Analysis (RCA) refers to a group of methods which are used to identify the root causes of an incident\citep{Andersen2006}. RCA is commonly used in a variety of industries, including security-critical domains like nuclear power \citep{iaeatecdoc596} and aviation \citep{IATA2016}.}  to identify why this incident occurred\footnote{The government agency may not deem it appropriate to perform a root cause analysis on every incident. It is outside the scope of this proposal to suggest criteria the government agency should follow in choosing whether or not to investigate an incident further.}. To aid with this root cause analysis, we suggest that the government agency should have the authority to gather `intelligence' from the AI provider. This could include access to documentation like the AI providers' pre-deployment national security case in order for the agency to identify why the incident happened (e.g., to determine whether the incident was a class of threat that the AI provider had not accounted for vs. a known threat that the provider's safety and security measures failed to counter). It could also involve the AI provider giving the government agency API or white-box access to the AI system involved in the incident in order to help the agency determine which inputs or tools were used to enhance the model's capabilities -- see  Section \ref{subsec:sketch} for more concrete examples of what access the government agency might require. 

Further, we suggest that, in light of the root cause analysis, the government agency should have the power to make some recommendations for how AI providers should change their security and safety procedures in order to counter similar threats in the future\footnote{The recommended changes to security and safety procedures should be proportionate to the threats involved. It is outside the scope of this paper to give more detail on how this should be adjudicated, but we note this as a promising area for further research.}. More concretely, we propose that the government agency should have the `security-strengthening authority,' under which AI providers are legally required to comply with these security and safety recommendations\footnote{Even if it performs a root cause analysis into the incidents, the government agency may not deem it appropriate to announce recommendations on how security and safety procedures should be improved in light of the incident. It is outside the scope of this proposal to suggest criteria the government agency should follow in choosing whether or not to make or enforce recommendations on changes to security and safety procedures.}. For example, this could involve directing AI providers to amend their national security cases in order to take into account a new threat vector that was not previously known about (perhaps the use of AI systems to perform sophisticated spear-phishing campaigns) -- again, see Section \ref{subsec:sketch} for more concrete examples of what kinds of recommendations the government agency could make. 

Next, we detail how government agencies that oversee other security-critical sectors have close equivalents to `intelligence-gathering' and `security-strengthening' authorities, wherein they amended rules on safety procedures following particular incidents. First we consider nuclear power, where the NRC has the authority to investigate nuclear utilities to work out the cause(s) of major incidents \citep{nrcmd83}, and to change the rules on safety procedures as a result. As part of this, nuclear investigators are explicitly instructed to use root cause analysis, taking actions like 1-1 interviews with key staff, reading the plant operating logs, and reading descriptions of safety procedures and equipment  \citep{iaeatecdoc596}. These investigatory powers support the NRC's rule-changing authority, which it has used in direct response to multiple incidents. For example, one intervention was that following a devastating fire at the Browns Ferry nuclear power plant in 1975, NRC started a fire safety research program, which in 1980 led to new rules around plants' fire safety systems. More notable was the NRC's response to the 1979 Three Mile Island incident, after which it rolled out a resident inspector program, wherein every nuclear plant had at least two NRC staff working on-site at all times, and switched to a new safety approach which involved calculating explicit probabilities that various adverse events occurred \citep{wellock2014}. In life sciences DURC, FSAP has the authority to access, in-person and without warning, any lab working with certain dangerous organisms in order to evaluate its safety procedures and inspect lab records \citep{cfr7318}. Finally, in aviation, the federal agency National Transportation Safety Board (NTSB) is empowered to investigate the causes of major incidents (\citep{usc1131}) by traveling to the site of the accident, with full access to flight logs/maintenance records and the power to interview relevant staff \citep{ntsb2024}. The NTSB does not have regulatory powers, so following an incident,  the FAA issues new safety rules that must be implemented by aviation companies (see, e.g., \citep{faa2002}). A particularly noteworthy example is the FAA introducing rules mandating reinforced cockpit doors \citep{cfr129, faa2002} due to the 9/11 attacks. Another example is the FAA introducing stringent new requirements for second-in-command pilots in direct response to the Colgan flight 3407 \citep{faa2013}, with the FAA raising the minimum age for the second-in-command to 23 and the minimum flight time from 190 hours to 1500 hours.

This Section proposed that an AI incident regime should have a `Hardening Defenses Phase,’ wherein a relevant government agency has both: (i) `intelligence-gathering' authority to access all necessary information required to perform root cause analysis (if they it deem appropriate) and (ii) `security-strengthening' authority such that if they recommend changes to AI providers' security and safety measures, AI providers must implement these changes. We justified these aspects of the AI incident regime's `Hardening Defenses Phase’ by demonstrating similarity to existing incident regimes in other security-critical sectors. 

\section {Sketch of the lifecycle of our AI incident regime proposal}
\label{subsec:sketch}

In the previous sections, we detailed three core phases that together make up our proposal for AI incident regime focussed on threats to national security -- a `Preparatory Phase', a `Rapid Response Phase' and then a `Defense-Hardening Phase'. In presenting those, we drew on evidence and case studies from other security-critical sectors, in particular nuclear power, aviation, and life sciences DURC. We now present a hypothetical example of how our proposed AI incident regime would function against a spear-phishing incident on the electricity grid, briefly articulating each step and putting it into (hypothetical) action.

\paragraph{Preparatory Phase - national security case --} An AI provider has compiled a `national security case' ahead of the public release of its AI system.

\paragraph{Preparatory Phase - discovery and denotation of an incident --} An AI provider discovers that a malicious actor has used one of its frontier AI systems through the API to develop a sophisticated spear-phishing attack on employees at the electricity grid \citep{hazell2023}, with the presumed goal of taking the grid down. The AI provider analyses API calls to try to better understand the situation and reviews its national security case for the relevant AI system. The security case made the following claim: even though the AI system has the capability to help execute a cyber attack, the AI system's guardrails are robust against jailbreaking, and hence, it does not pose a cyber threat. The current situation weakens this claim, and so the AI provider determines that the spear-phishing event is an `incident.'

\paragraph{Rapid Response Phase - notification --} 6 hours after discovering that its AI system has been used to develop a sophisticated spear-phishing attack on the electricity grid, the AI provider notifies the relevant government agency through an emergency phone line. The notification contains several pieces of information gleaned during these 6 hours. It includes a brief description of the spear-phishing attack, mitigation actions taken by the AI provider (e.g., preventing API access from any devices with IP addresses associated with the malicious actor), and information about the identity of the malicious actor and their plans.

\paragraph{Response Phase - notification --} Once notified, the government agency coordinates with the electricity grid to warn them of the attacks, verify that the attack has not yet been executed, and ensure that grid employees are protected from future phishing attacks. They also share details about the malicious actor (e.g., IP addresses) to relevant law enforcement bodies. 
\paragraph{Response Phase - containment --} Once notified, the government agency coordinates with the electricity grid to warn them of the attacks, verify that the attack has not yet been executed, and ensure that grid employees are protected from future phishing attacks. They also share details about the malicious actor (e.g., IP addresses) to relevant law enforcement bodies. 

\paragraph{Defense-Hardening Phase - intelligence --} 20 days after the incident was first reported, and now that the incident has been contained, the government agency turns to the next steps. Given that the incident directly undermines a core claim in the AI system's national security case, they decide that this incident warrants proactive engagement -- they will perform a root cause analysis on the incident and, in light of this, may recommend changes to AI providers' safety and security countermeasures. In doing so, the agency exerts its intelligence-gathering authority and requests full cooperation and disclosure from the AI provider in question\footnote{We note that AI providers may collaborate throughout these investigations and provide valuable additional insight, given their technical expertise and deep familiarity with the relevant AI system.}. As part of this, it receives access to the complete log of AI system inputs and outputs used by the malicious actor leading up to the incident (e.g., log of API requests made by the malicious actor, the developer message, instances of the tool, the complete Chain of Thought). During the investigation, they identify that the malicious actor has used a new jailbreak to bypass Llamaguard, the AI system's classifier guardrails \citep{inan2023}. Next, they investigate why the safety/security systems failed. With access to the AI system API and logits \citep{goodfellow2016}, they rapidly iterate through system inputs to identify which parts of the input were important for the malicious actor jailbreaking the AI system \citep{greenblatt2024}. This completes their causal analysis.

\paragraph{Defense-Hardening Phase - security-strengthening --} Once the causal analysis is complete, the agency may recommend amendments to existing procedures based on what was learned from the analysis. In this hypothetical case, the agency recommends adjustments to the (i) governance measures, such as AI providers' Frontier AI Safety Policy (a high-level safety and security procedure)\footnote{Frontier AI Safety Policies are `if-then commitments for AI risk reduction' \citep{karnofsky2024, metr2024}.}, and (ii) technical defensive countermeasures, such as classifiers.

\textit{Adjustment to governance measures (i)} The provider's `national security case' had not foreseen the risk of spear-phishing attacks, and as such, the AI provider had not been evaluating the AI system's spear-phishing capabilities or explicitly attempting to mitigate these capabilities. The agency uses its `security-strengthening authority' to mandate that the `national security case' of this AI provider — and to the other AI providers that do not evaluate spear-phishing capabilities in their `national security cases'—be amended to take into account this new, unforeseen threat vector going forward. 

\textit{Adjustment to technical measures (ii)} The agency recommends changes to technical security and safety measures to make sure that systems with similar vulnerabilities are not deployed publicly in the future. Specifically, the classifier in this hypothetical scenario was Llama Guard, a small LLM that runs on top of the main model, checking whether user inputs are safe before allowing the LLM to respond and checking whether the AI output is safe before allowing it to reach the user \citep{inan2023}. Following its root cause analysis, the agency generates evaluation tasks by finding new inputs that would also have bypassed Llama Guard and allowed spear-phishing to occur. They speed up this process with their access to the API and logits, while access to the model's gradients allows the agency to perform white-box adversarial attacks\citep{goodfellow2015}\footnote{We note that reflections on whether and how a government agency will have the technical capacity to conduct these investigations is outside the scope of this paper.}. As such, the agency mandates that, notwithstanding other security and safety procedures, AI providers must perform pre-deployment jailbreak evaluations against a new set of evaluation test tasks it has created using data from Llama Guard in this incident. 

 As a result of this, similar spear-phishing attacks on critical national infrastructure are prevented from occurring in the future. 

\section{Conclusion}

We have put forward and justified a proposal for an AI incident regime that can help counter potential post-public deployment national security threats from AI systems. We started the paper by coining the term `security-critical' to describe sectors that pose extreme risks to national security and argued that 
`security-critical' describes frontier AI development, civilian nuclear power, aviation, and life sciences dual-use research of concern. Then, the main part of the paper detailed our proposal for an AI incident regime. We justified each component of our AI incident proposal by demonstrating that it mirrors U.S. incident regimes in other security-critical sectors. Our proposal had three phases: first, a `Preparatory Phrase' wherein, prior to public deployment,  AI providers must create national security cases for their frontier AI systems before they are deployed, and after which they must report as an incident any event that weakens a claim made in a `national security case' for the AI system. Secondly, the `Rapid Response Phase': providers must notify a government agency no later than 24 hours after discovering an incident has occurred so that, if appropriate, a the government agency can coordinate an emergency containment response. And finally, the `Hardening Defenses Phase' involves countering future threats to national security. Here, we recommend that the government agency have two authorities -- an `intelligence-gathering authority' which gives the government agency sufficient access to investigate the incident, and a `security-strengthening authority' which allows the government agency to direct all relevant AI providers to amend their security and safety procedures. Against a backdrop where AI systems' ability to pose national security threats has rapidly progressed and may continue to increase, the three phases of our proposal are grounded in common practice in other security-critical sectors, offering an intervention that can help to track and counter the national security threats that may be posed by AI systems. 

\bibliography{example_paper}
\bibliographystyle{icml2025}

\end{document}